\documentclass[usenatbib,final]{mn2e}
\usepackage{graphicx,graphbox,tabularx,amsmath,amssymb,upgreek,wasysym,mathtools,multirow,url,rotating,pdflscape,lastpage,stfloats}
\numberwithin{equation}{section}

\usepackage[backgroundcolor=blue!10!white,obeyFinal]{todonotes}
\graphicspath{{figures/}}

\newcommand{\Jo}{J_{1}}
\newcommand{\Jt}{J_{2}}

\title[$J$ plots]{$J$ plots: a new method for characterizing structures in the interstellar medium}
\author[S. E. Jaffa, A. P. Whitworth, S. D. Clarke \& A. D. P. Howard]{S. E. Jaffa$^{1}$\thanks{E-mail: Sarah.Jaffa@astro.cf.ac.uk}, A. P. Whitworth$^{1}$, S. D. Clarke$^{2}$ and A. D. P. Howard$^{1}$\\  
$^{1}$School of Physics and Astronomy, Cardiff University, Cardiff CF24 3AA, Wales, UK\\ 
$^{2}$I. Physikalisches Institut, Universit{\"a}t zu K{\"o}ln, Z{\"u}lpicher Str. 77, D-50937 K{\"o}ln, Germany}

\begin{document}
\newpage
\pagerange{\pageref{firstpage}--\pageref{lastpage}} \pubyear{2018}
\maketitle
\label{firstpage}

\begin{abstract} 
Large scale surveys have brought about a revolution in astronomy. To analyse the resulting wealth of data, we need automated tools to identify, classify, and quantify the important underlying structures. We present here a method for classifying and quantifying a pixelated structure, based on its principal moments of inertia. The method enables us to automatically detect, and objectively compare, centrally condensed cores, elongated filaments and hollow rings. We illustrate the method by applying it to (i) observations of surface-density from Hi-GAL, and (ii) simulations of filament growth in a turbulent medium. We limit the discussion here to 2D data; in a future paper we will extend the method to 3D data.
\end{abstract}

\begin{keywords}
Stars: formation, ISM:structure
\end{keywords}

\section{Introduction}%

The multi-phase interstellar medium (ISM) is a chaotic environment in which many physical processes interact over a large range of scales. These processes sculpt the ISM into a variety of structures, and many of these structures play important roles in determining and/or revealing the locations of star formation. For example, \textit{Herschel} observations have shown that elongated filamentary structures are a common feature in molecular clouds, in both low- and high-mass star forming regions, and also in regions that are not actively forming stars \citep{2010Molinari+, 2010Andre+, 2012Hennemann+, 2013Palmeirim+}. Further studies have estimated that 60\% to 75\% of bound prestellar cores lie on filaments whose line mass is greater than the critical value for the onset of fragmentation \citep{2010Andre+, 2015Konyves+, 2016Marsh+}. Feedback processes, from small scale stellar winds up to individual or clustered supernovae, can create hollow shells or bubbles which show up in observations as dense rings, sometimes with an ionising source at the centre \citep{2006Churchwell+, 2012Simpson+, 2016Ingallinera+}. Finally, gravitationally bound objects, from clouds to cores and even clusters of stars, are often characterised as being fractally substructured and/or centrally condensed \citep[\textit{subm.}]{2004Cartwright+, 2015Li+, 2015Rathborne+, 2016Storm+,2017Jaffa+,2017Sills+}. Each of these different types of structure can coexist in astrophysical images, and are relevant to different parts of the star formation process. We therefore need methods to identify and quantify such structures, in order to compare their properties and build a coherent picture of the interaction and interdependence of stars and the ISM.

\todo[inline]{Update Sills reference if poss}

A variety of algorithms is used in star formation to segment astronomical images into regions of interest. Some are designed to find particular shapes, such as filaments \citep{2011Sousbie, 2013Menshchikov, 2014Schisano+,2017Koch+} or rings \citep{2012Simpson+}. Others use intensity or surface-density information to group pixels with similar properties into larger structures \citep{2008Rosolowsky+, 2015Li+, 2015Colombo+}. Compact sources can be extracted \citep{2011Molinari+, 2012Menshchikov+} and their distribution analysed \citep{2015Parker+, 2017Jaffa+, 2017Joncour+}. Statistical techniques such as Principal Component Analysis can be used to examine large datasets and mathematically define groupings in the data, thereby constraining the underlying physics \citep{2017Gratier+}. In this work we use dendrograms to segment images into an hierarchical set of contiguous structures, using the \textsc{astrodendro}\footnote{http://www.dendrograms.org/} Python package. We then analyse these structures using their principal moments of inertia. We limit the discussion to the case of 2D data; in a future paper we will extend the method to 3D data, using it (a) to explore the relationship between structures identified in PPP and PPV data-cubes from simulations, and (b) to inform the interpretation of structures identified only in PPV data-cubes from observations.

In Section \ref{Methods} we briefly describe the use of dendrograms for segmenting astronomical images into discrete structures;  we explain how the principal moments of inertia can be combined to construct $J$ moments, which distinguish structures according to their degree of central concentration (or central rarefaction) and their degree of elongation; and we illustrate the use of $J$ plots to quantify some simple synthetic images. In Section \ref{Results} we give examples of the application of $J$ plots to observed and simulated data. In Section \ref{SEC:Conclusions} we summarise our main conclusions.

\begin{table*}
\begin{center}
\caption{Some simple structures, and the constraints on their principal moments ($I_1,I_2$) and $J$ moments ($J_1,J_2$).}
\begin{tabular}{lccc}
{\sc Structure} & {\sc Principal Moments} & $J_1$ & $J_2$ \\
Uniform surface-density disc & $I_1=I_2=I_0\equiv AM/4\pi$ & $0$ & $0$ \\
Centrally concentrated disc (core) & $I_1=I_2<I_0\equiv AM/4\pi$ & $>0$ & $>0$ \\
Ring  (limb-brightened bubble) & $I_1=I_2>I_0\equiv AM/4\pi$ & $<0$ & $<0$ \\
Elongated ellipse (filament) & $I_1<I_0\equiv AM/4\pi<I_2$ & $>0$ & $<0$ \\
\end{tabular}
\end{center}
\label{TAB:MOMENTS}
\end{table*}

\section{Methods}\label{Methods}%

In this section we give a brief outline of how dendrograms are used to segment a pixelated 2D grey-scale image into hierarchically nested structures. Then we describe how these structures can be analysed using $J$ moments and $J$ plots. In order to make the discussion less abstract, we assume that the intensity of the 2D grey-scale image being analysed represents surface-density, $\Sigma$. Hence the grey-scale contours are surface-density thresholds. For a more comprehensive treatment of dendrograms, see \citet{2008Rosolowsky+}.

\subsection{Image segmentation with dendrograms}\label{Image seg}

Dendrograms represent the morphology of a greyscale image in terms of hierarchically nested contiguous structures at different surface-density thresholds. At the specified minimum surface-density threshold (`min\_value' parameter in \textsc{astrodendro}), there is usually a single contiguous structure, which is termed the `trunk' of the tree. As the surface-density threshold is increased, the trunk splits into smaller `branch' structures.\footnote{The \textsc{astrodendro} package imposes binary mergers, meaning that if a larger structure at surface-density threshold $i$ splits into 3 smaller structures at surface-density threshold $i + \delta i$, additional intermediate thresholds are examined to find one at which the large structure splits into two structures, one of which will split again. Other dendrogram building methods such as that described in \citet{2014Storm+} do not enforce binary mergers.} A structure that does not split into smaller structures but reaches its peak surface-density as a contiguous whole is termed a `leaf', and corresponds to a local maximum. Leaves are only retained if (i) they comprise a minimum number of pixels (`min\_pixels' in \textsc{astrodendro}) and (ii) there is a minimum surface-density contrast between their peak surface-density and the surface-density threshold at which they merge with another leaf or branch (`min\_delta' in \textsc{astrodendro}). Leaves that do not satisfy both these conditions are simply absorbed into a larger structure. Thus a dendrogram analysis divides any greyscale image into a set of structures defined by their surface-density thresholds, and results in a two dimensional `tree' graph which summarises the connectivity of structures at different thresholds. In this paper we extract the structures defined by the dendrogram, and use moments to measure their degree of central concentration and their overall shape.

We must chose carefully the parameters of the dendrogram, so as to capture the structures we are interested in. Reducing the minimum surface-density threshold (`min\_value') will increase the size of the trunk, until it encompasses the entire observed area. Conversely, increasing the minimum threshold may split the dendrogram into separate trees, i.e. multiple trunks. The minimum number of pixels (`min\_npix') and minimum surface-density contrast (`min\_delta') can change what is defined as a leaf. Reducing either of these parameters results in more leaves. The intermediate branches of the tree are little affected by these parameters.

\begin{figure*}
\includegraphics[width=1.0\linewidth]{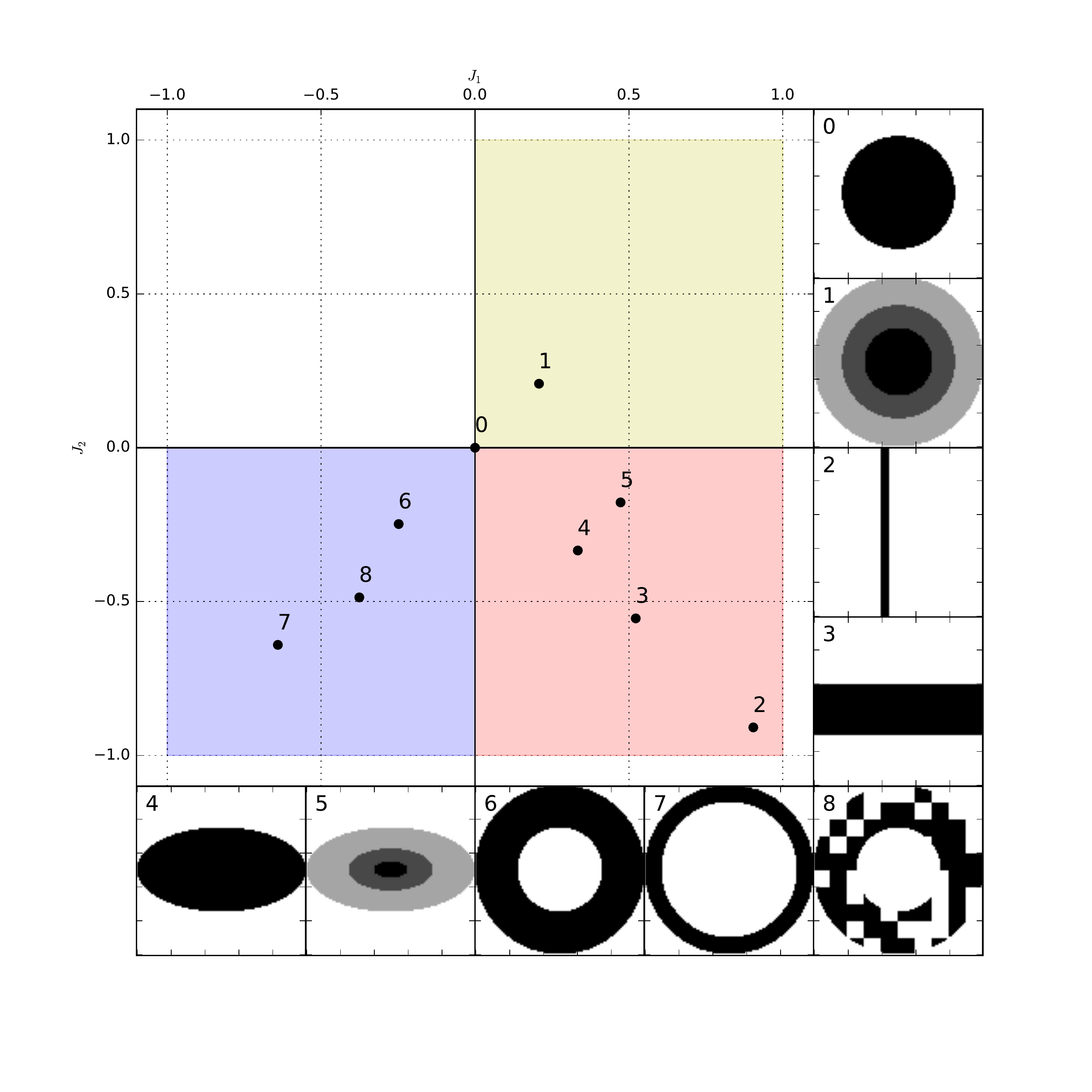}
\caption{Some simple two dimensional test-structures and their positions on the $J$ plot. A circularly symmetric disc with uniform surface-density (test-structure 0) occupies the centre of the plot, $J_1=J_2=0$. Circularly symmetric, centrally concentrated discs (test-structure 1) occupy the top right quadrant, getting closer to $J_1=J_2=1$ as their central concentration increases. Circularly symmetric, hollow structures (test-structures 6, 7 and 8) occupy the bottom left quadrant, getting closer to $J_1=J_2=-1$ as their thickness decreases. Elongated objects (test-structures 2, 3, 4 and 5) occupy the bottom right quadrant, getting closer to $(J_1,J_2)=(1,-1)$ as their aspect ratio increases. Nothing falls in the top left quadrant, as this would require $I_1 > I_2$.}
\label{fig:proof}
\end{figure*}

\subsection{$J$ moments}\label{Moments}

Any point on a dendrogram (be it on a leaf, branch or trunk) corresponds to a contiguous region ${\cal R}$ inside which the surface-density, $\Sigma(x,y)$, exceeds or equals some threshold, $\Sigma_{_{\cal R}}$. We can therefore compute the total area, $A$, and mass, $M$, of the corresponding object, 
\begin{eqnarray}
A&=&\int\limits_{\cal R}\,dx\,dy\hspace{1.2cm}=\;\,P\;\Delta A\,,\\
M&=&\int\limits_{\cal R}\,\Sigma(x,y)\,dx\,dy\;\,=\;\,\sum\limits_{p=1}^{p=P}\,\left\{\Sigma_p\right\}\;\Delta A\,.
\end{eqnarray}
Here $p$ is the dummy ID of a pixel inside ${\cal R}$, $P$ is the total number of pixels inside ${\cal R}$, $\Sigma_p$ is the surface-density in pixel $p$, and $\Delta A$ is the area of a pixel; we are assuming that all pixels have the same area.

We can also compute the principal axes of this object, $\hat{\bf e}_i\;\;(i=1,2)$, and the associated principal moments, $I_i$ (see Appendix). By convention, if the two principal moments are different (which is generally the case), the first principal axis, $\hat{\bf e}_1$ is the one associated with the smaller principal moment. Hence $I_1\leq I_2$, by construction.

Now consider the very simple case of an infinitesimally thin circular region (a disc) with radius $R$, and uniform surface-density $\Sigma_{_{\rm O}}$, hence area $A=\pi R^2$ and mass $M=\pi R^2\Sigma_{_{\rm O}}$. Because of symmetry, the principal axes can be any pair of orthogonal axes, and without loss of generality we choose the Cartesian axes, i.e. $\hat{\bf e}_1\rightarrow\hat{\bf e}_x$ and $\hat{\bf e}_2\rightarrow\hat{\bf e}_y$. The corresponding principal moments are
\begin{eqnarray}
I_1\!\!&\!\!=\!\!&\!\!I_2\;=\;\frac{1}{2}\int\limits_{r=0}^{r=R}r^2\,\Sigma_{_{\rm O}}\,2\pi rdr\;=\;\frac{\pi R^4\Sigma_{_{\rm O}}}{4}\;=\;\frac{AM}{4\pi},
\end{eqnarray}
where we have invoked the Perpendicular Axis Theorem.

Next, whilst maintaining circular symmetry, change the radial profile of the surface-density, $\Sigma(r)$, holding the area, $A$, and mass, $M$, constant. If the surface-density profile is made more centrally concentrated (e.g. a centrally condensed 3D core seen in projection), $I_1$ and $I_2$ are decreased, and therefore $I_1=I_2<AM/4\pi$. Conversely, if the surface-density profile is altered so it has a central rarefaction (e.g. a ring or 3D limb-brightened bubble), $I_1$ and $I_2$ are increased, and therefore $I_1=I_2>AM/4\pi$.

Finally, revert to a uniform surface-density, and again hold the area, $A$, and mass, $M$, constant, but stretch the disc into a very eccentric ellipse (i.e. a filament). This has the effect of reducing $I_1$ and increasing $I_2$, so we have $I_1<AM/4\pi <I_2$.

Now, if we introduce
\begin{eqnarray}
I_0&=&\frac{AM}{4\pi}\,,
\end{eqnarray}
and define the $J$ moments
\begin{eqnarray}\label{EQN:JMoments}
J_i&=&\frac{I_0-I_i}{I_0+I_i}\,,\hspace{0.8cm}i=1,2\,,
\end{eqnarray}
the uniform surface-density disc has $J_1=J_2=0$; the centrally concentrated disc (3D core in projection) has $J_1=J_2 >0$; the centrally rarefied disc (ring or 3D limb-brightened bubble) has $J_1=J_2<0$; and the eccentric ellipse (filament) has $J_1>0,\;J_2<0$. Thus, if we define a $J$ space with abscissa $J_1$ and ordinate $J_2$, centrally concentrated objects occupy the top right quadrant, centrally rarefied objects occupy the bottom left quadrant, and elongated objects occupy the bottom right quadrant. These results are summarised in Table \ref{TAB:MOMENTS}.

\begin{table}
\centering
\caption{Properties of the test-structures}
\begin{tabular}{lllrr}
{\sc Test Structure} & {\sc Area} & {\sc Mass}  & $\Jo\;\,$    & $\Jt\;\,$    \\
    & (pixels) & (arbitrary units) & & \\
1     & 3505 & 3505  & 0.00  & 0.00  \\
2     & 7825 & 17325 & 0.21  & 0.21  \\
3     & 500  & 500   & 0.90  & -0.91 \\
4     & 3000 & 3000  & 0.52  & -0.55 \\
5     & 3895 & 3895  & 0.33  & -0.33 \\
6     & 3895 & 6135  & 0.47  & -0.18 \\
7     & 5884 & 5884  & -0.25 & -0.25 \\
8     & 2812 & 2812  & -0.64 & -0.64 \\
9     & 2976 & 2976  & -0.51 & -0.53
\end{tabular}
\label{TAB:TestImages}
\end{table}

\begin{figure}
\centering
\hspace{-0.4cm}\includegraphics[width = \linewidth]{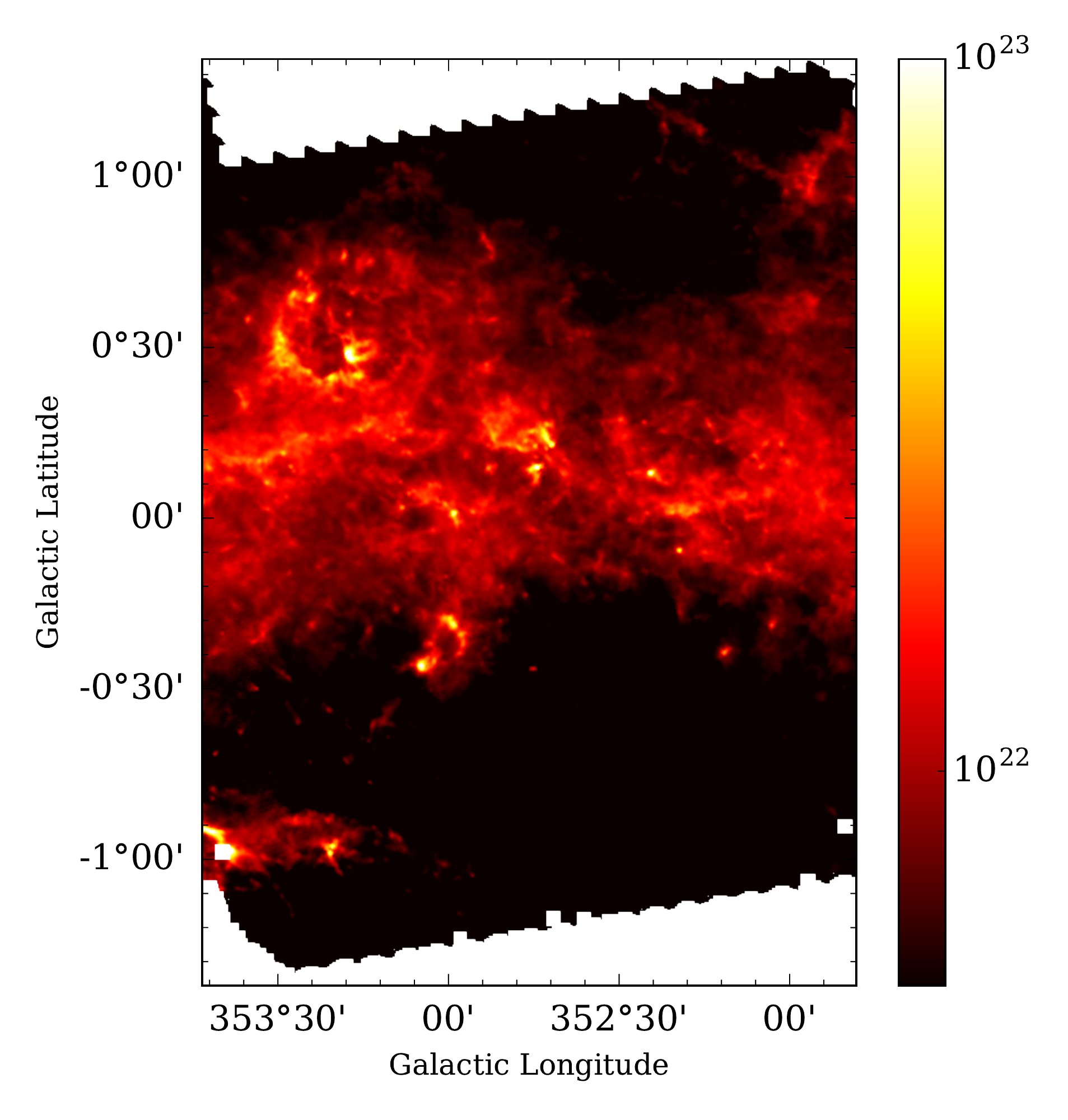}\hspace{0.2cm}
\caption{Hi-GAL surface-density tile $\ell$347. The colour bar gives surface-density, on a logarithmic scale, and in units of ${\rm H}_2\,{\rm cm}^{-2}$.}
\label{fig:rcw120}
\end{figure}

\begin{figure}
\centering
\hspace{0.60cm}\includegraphics[width = 0.87\linewidth]{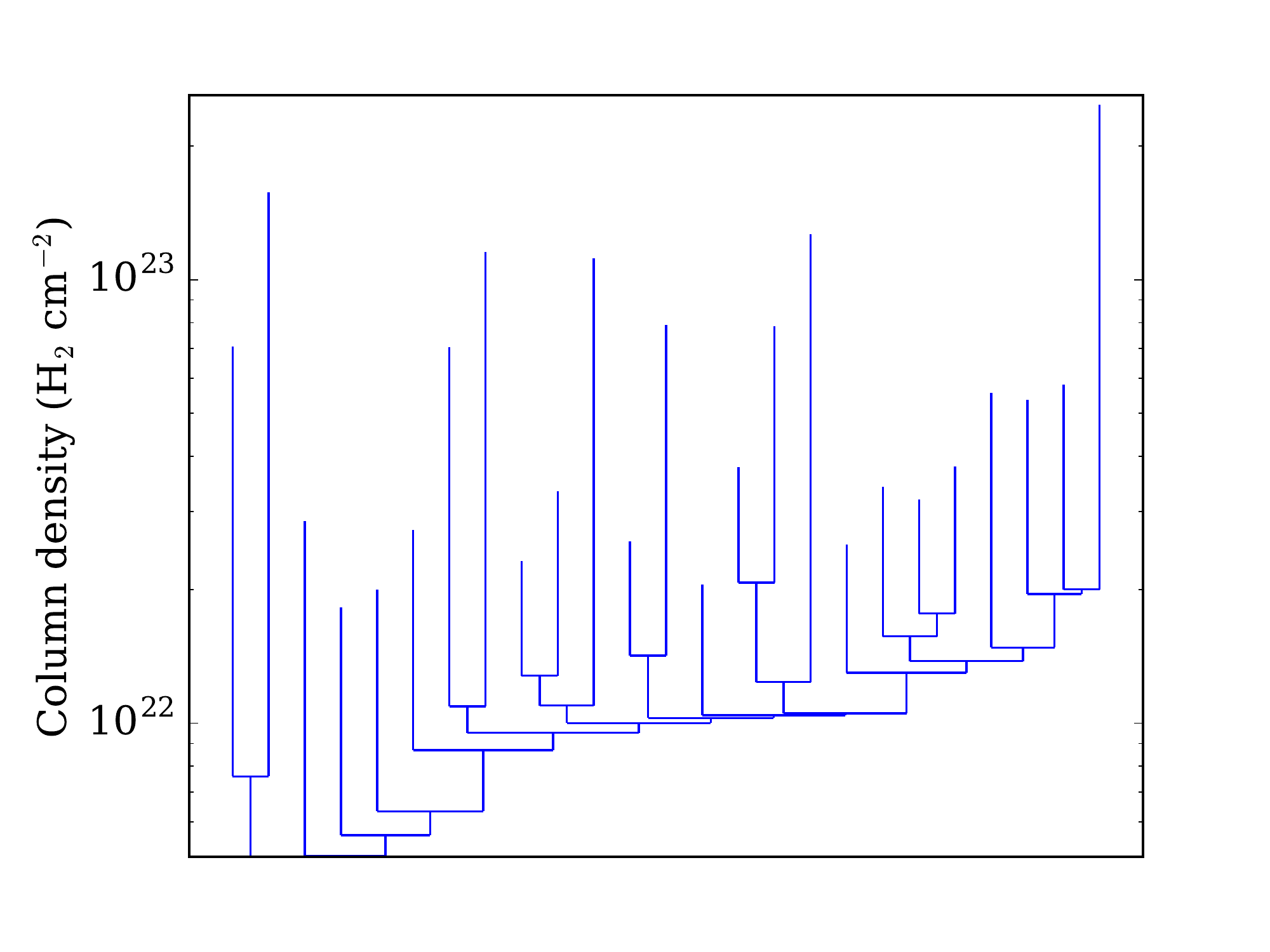}\\
\includegraphics[width = 0.9\linewidth]{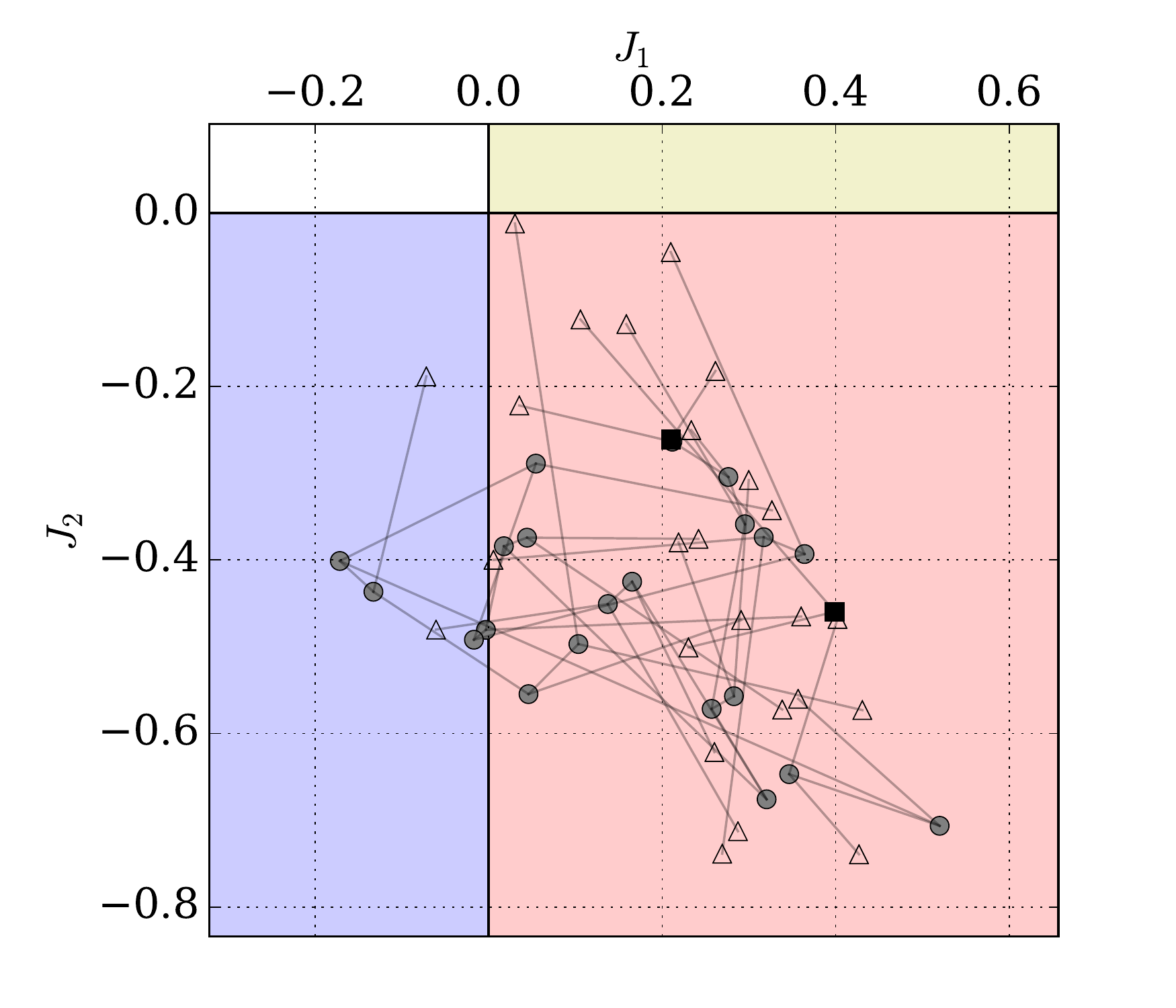}
\caption{{\it Top:} the dendrogram of the HI-GAL tile shown in Figure \ref{fig:rcw120} constructed using the parameters specified in the text. {\it Bottom:} the $J$ plot of the identified structures. Leaves are shown as triangles, branch structures are shown as circles, and the trunks are shown as squares. Grey lines show which structures merge to form larger structures. Note that this image is focused on only the portion of the $J$ plot occupied by the detected structures. To aid orientation, the background colouring is identical to that in Figure \ref{fig:proof}.}
\label{fig:rcw120dend}
\end{figure}

\subsection{$J$ plots}\label{J plots}

Figure \ref{fig:proof} shows a $J$ plot (i.e. a plot with $J_1$ on the abscissa and $J_2$ on the ordinate); filled circles mark the locations of the simple test-structures that are illustrated -- with pixelated images -- in the margin. For each test-structure, the corresponding area, mass and $J$ moments are given in Table \ref{TAB:TestImages}. The circle with uniform surface-density (test-structure 0) occupies the origin ($\Jo=\Jt=0$). The circularly symmetric, centrally concentrated disc (test-structure 1) occupies the upper right quadrant ($\Jo =\Jt > 0$) as the principal moments are equal, $I_1=I_2$, but both less than $I_0$; such discs move further from the origin with increasing central concentration. The circularly symmetric hollow rings (test-structures 6 and 7) occupy the lower left quadrant ($\Jo =\Jt <0$) as their principal moments are equal, $I_1=I_2$, but both greater than $I_0$; such rings move further from the origin with decreasing thickness (decreasing $\Delta R/R$). Elongated straight structures with uniform surface-density (test-structures 2, 3, and 4) occupy the lower right quadrant ($\Jt < 0 < \Jo$) as $I_1<I_0$ and $I_2>I_0$; elongated structures move further from the origin with increasing aspect ratio. Nothing falls in the top left quadrant, or indeed above the line $J_2=J_1$, since  this would imply $I_{2} < I_{1}$, in violation of the convention that $I_1 \leq I_2$.

Structures that do not fit perfectly into one of these patterns can still be represented usefully on the $J$ plot. For example, the broken ring (test-structure 8) still falls in the lower left quadrant of the $J$ plot. And the centrally concentrated ellipse (test-structure 5) falls between the locus for elongated structures with uniform surface-density and the locus for circularly symmetric, centrally concentrated structures, indicating that it has elements of both.

A single dendrogram structure (leaf, branch or trunk) spans a range of surface-densities, and in general its $J$ moments will depend on the surface-density at which we analyse it. For the test-structures in Figure \ref{fig:proof}  we have treated each structure in its entirety, which is equivalent to analysing it at its base, i.e. at the lowest surface-density before it merges with another structure. However, we can also look at how the $J$ moments of a dendrogram structure migrate on the $J$ plot as the surface-density threshold is changed. For example, the circularly symmetric, centrally concentrated test-structure on Figure \ref{fig:proof} (test-structure 1) would become less centrally concentrated and migrate towards the origin if we were to analyse it at higher and higher surface-density thresholds; the mapping of surface-density onto this path would depend on the details of the surface-density profile.

\section{Applications}\label{Results}
\subsection{Bubbles in observations}

The \textit{Herschel} Hi-GAL survey has delivered an unprecedented view of the Galactic plane in five wavebands between $70\,\mu{\rm m}$ and $500\,\mu{\rm m}$ \citep{2016Molinari+}, and the images in the different wavebands have been smoothed to a common resolution and fit with SEDs to obtain surface-density maps. As a simple example of the application of $J$ plots to observational images, we take the `tile' from the Hi-GAL survey shown in Figure \ref{fig:rcw120}, which contains the well known RCW 120 bubble centred near $\ell = 353.20^{\rm o},\;b = 0.30^{\rm o}$. This HII region has been described as `the perfect bubble' \citep{2009Deharveng+}, and much work has been done studying the physical conditions in and around the bubble \citep{2010Anderson+, 2013Pavlyuchenkov+, 2015Rodon+}, and the likelihood of triggered star formation in the swept up shell \citep{2010Motte+, 2015Walsh+, 2017Figueira+}. We therefore use this well characterized bubble as a test case to see how it appears on the $J$ plot.

To construct the dendrogram we use a minimum surface-density threshold (`min\_value') of $5 \times 10^{21} \mathrm{cm}^{-2}$, a minimum surface-density contrast (`min\_delta') of $1 \times 10^{22} \mathrm{cm}^{-2}$, and a minimum number of pixels (`min\_npix') of 200. These values are deliberately chosen to be quite high, since we are looking for large, clearly defined structures. The value of the minimum surface-density threshold results in two separate trunks. In total the dendrogram contains 48 structures (trunks, branches and leaves). 

\begin{figure*}
\includegraphics[width=\linewidth]{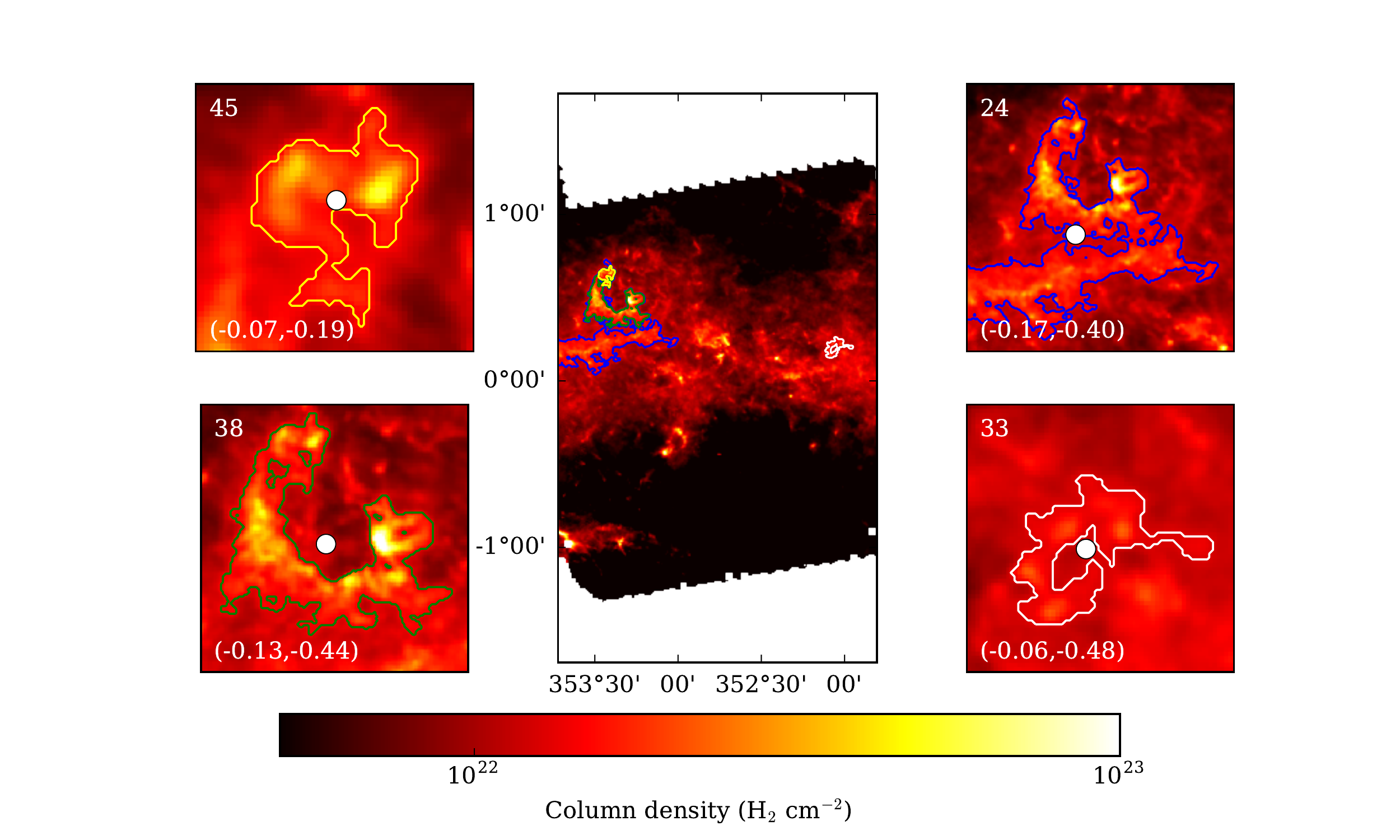}
\caption{The four structures from Hi-GAL tile $\ell$347 that are classified as rings, and their positions in the tile. In each panel, the structure ID is given in the top left corner, the $J$ moments in the bottom left corner, and the centre of mass is marked by a white circle. We note that structure 45 is embedded in structure 38, which in turn is embedded in structure 24.}
\label{fig:rcw120rings}
\end{figure*}

For each dendrogram structure, we consider the lowest surface-density threshold (i.e. the surface-density at which it merges with another structure), and compute $\Jo$ and $\Jt$. The bottom panel of Figure \ref{fig:rcw120dend} shows the $J$ plot for this data set, focusing on the region of the $J$ plot that is actually occupied. We use different symbols on the $J$ plot to distinguish different types of dendrogram structure: the highest level structures, leaves, are shown as triangles; branch structures are shown as circles; and the trunk is shown as a square. Lines show where smaller high-level structures merge to form a larger lower-level structure. Most of the structures fall in the lower right quadrant, indicating that they are elongated. Six structures fall in the lower left quadrant, indicating that they are `ring-like', although two are very close to the boundary and therefore not clearly defined as rings. The morphologies of the 4 structures clearly defined as rings are shown in Figure \ref{fig:rcw120rings}. We note that structure 45 is embedded in structure 38, which in turn is embedded in  structure 24.

Structure 38 is the region identified as the RCW 120 bubble; in this surface-density image it is a partial ring with a lumpy boundary. Structure 45 is a small leaf structure at the top edge of the bubble wall. A visual inspection confirms a broken ring type morphology, but its astrophysical significance is unclear. Structure 24 is formed by the merging of the RCW 120 ring structure with a filamentary structure to the south. This creates an asymmetrical curved shape, but a visual inspection would not readily identify this as a bubble candidate. Structure 33 is a collection of 4 cores that appear to be spaced regularly around a circle. These sources have been confirmed as compact sources by other surveys including the Hi-GAL point source catalogue \citep{2017Elia+} and the ATLASGAL dust condensation catalogue \citep{2014Csengeri+}. However their unusually regular arrangement has not been noted previously. We can speculate that they might be a set of fragments formed out of a swept up ring or shell, but equally they might be at different distances and causally unrelated.

As demonstrated in Figure \ref{fig:proof}, the distance from the origin in the bottom left quadrant is associated with the thickness of the ring, although this is only an exact relation in the case of idealised test-structures. The $J$ values, given on Figure \ref{fig:rcw120rings} for each of the ring-like structures, support this interpretation: the thickest ring (structure 45) lies significantly closer to the origin than the others, and the thinnest (structure 33) lies furthest from the origin.

\subsection{Filaments and cores in simulations}

\subsubsection{Sub-structure in filaments}

\begin{figure*}
\includegraphics[width=0.49\linewidth]{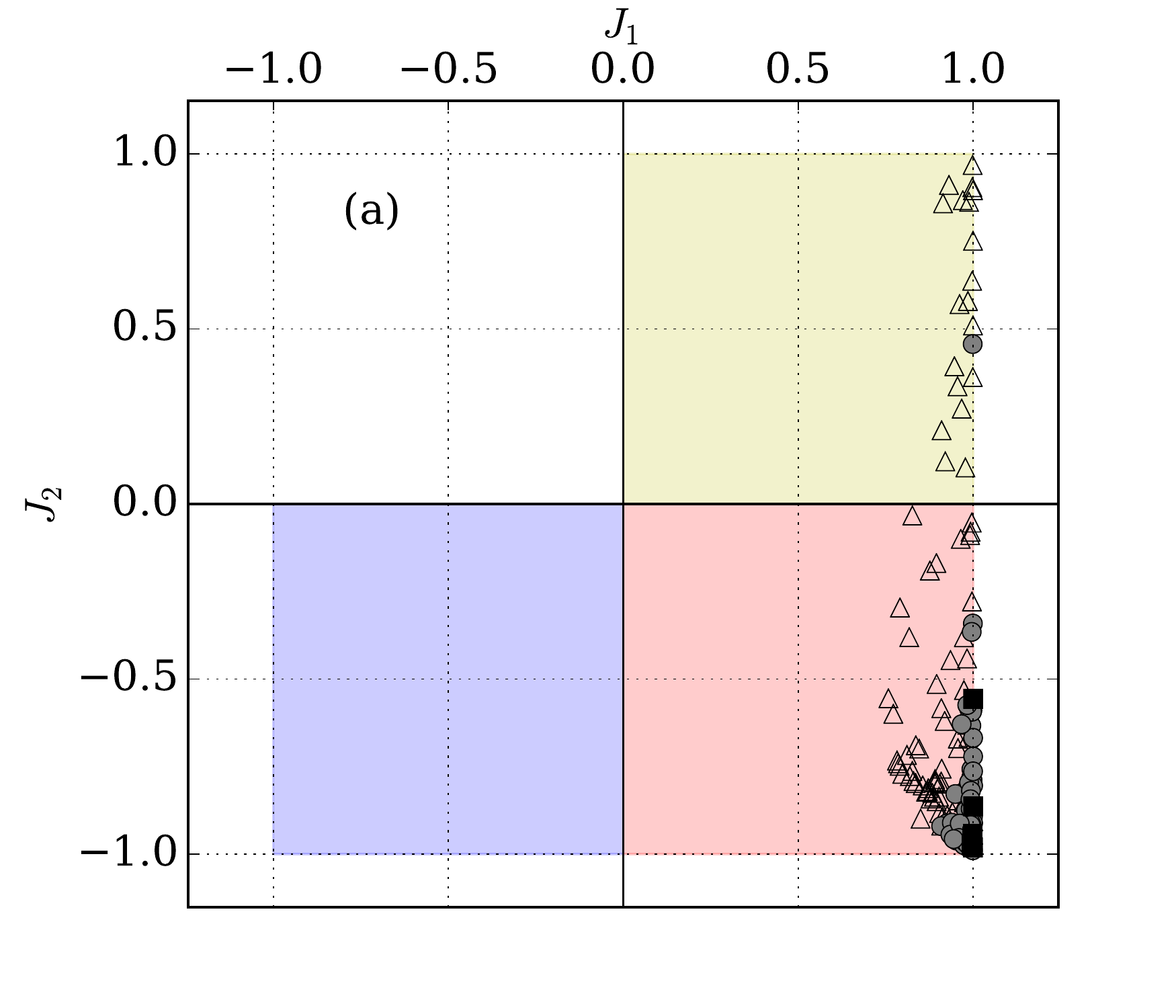}
\includegraphics[width=0.49\linewidth]{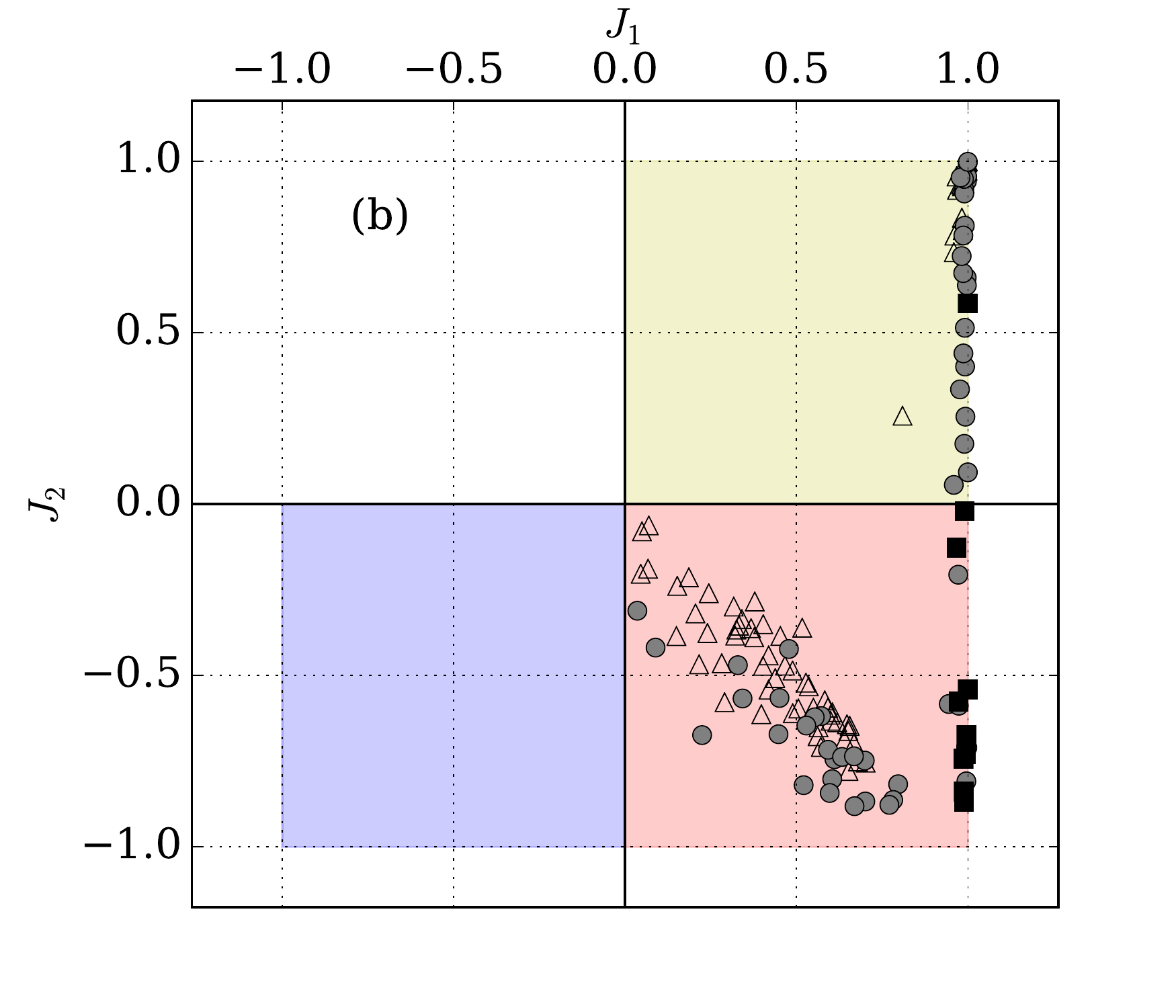}
\caption{$J$ plots of simulations of filament formation \citep{2017Clarke+} in which the turbulence in the accretion flow involves a thermal mix of modes, and is (a) subsonic or (b) supersonic. Each plot presents the results of 10 simulations, analysed at the end when 10\% of the mass has been accreted into sink particles. The open triangles are leaves, the grey circles are branches, and the black squares are trunks.}
\label{fig:filssubvsuper2D}
\end{figure*}

\begin{figure*}
\includegraphics[width=0.49\linewidth]{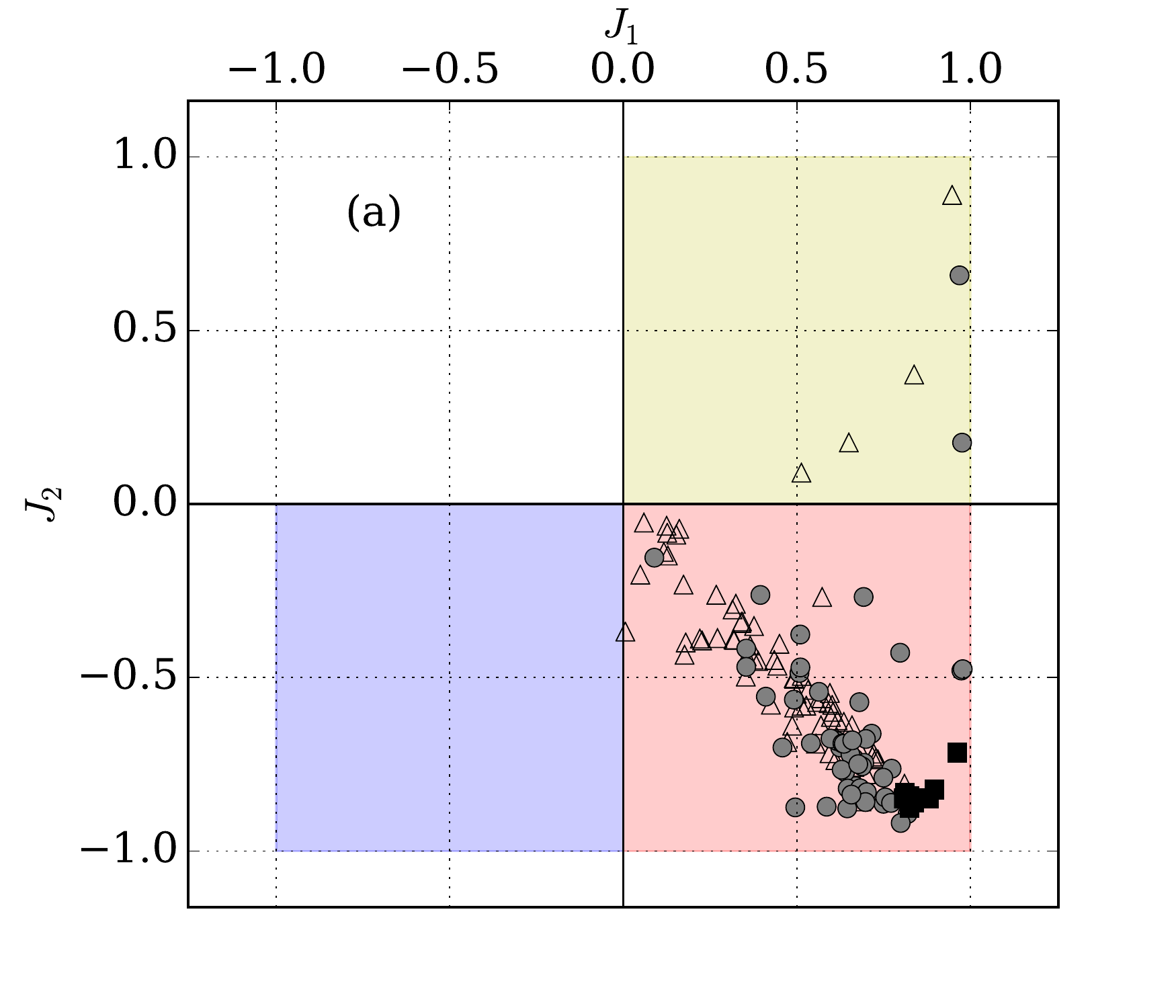}
\includegraphics[width=0.49\linewidth]{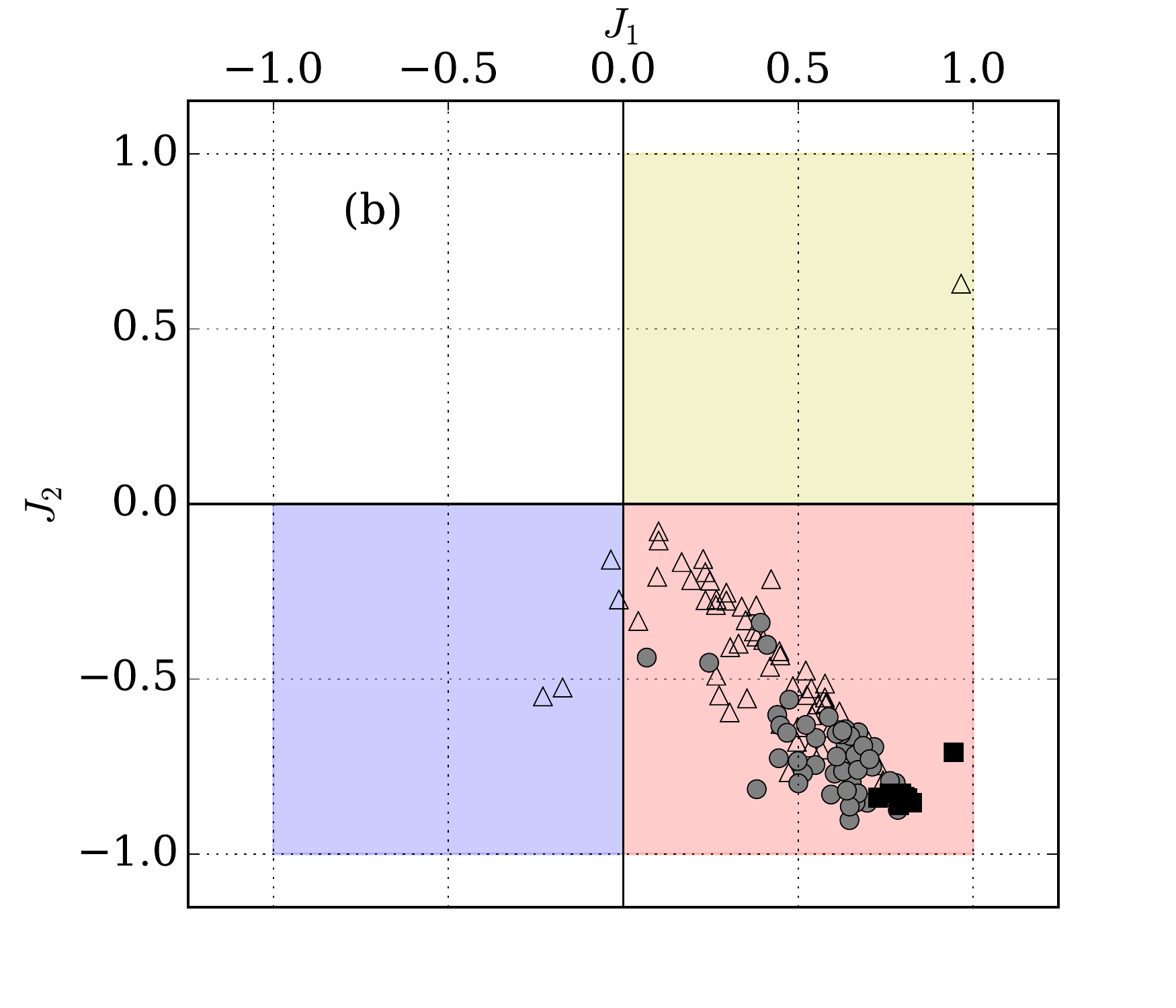}
\caption{$J$ plots for simulations of filament formation \citep{2017Clarke+} in which the turbulence in the accretion flow is supersonic, and involves (a) purely compressive modes, or (b) a thermal mix of modes. Each plot presents the results of 10 realisations, analysed after $0.5\,{\rm Myr}$. The open triangles are leaves, the grey circles are branches, and the black squares are trunks.}
\label{fig:filsnatvcomp2D}
\end{figure*}

\citet{2017Clarke+} have performed SPH simulations of filaments forming by accreting from a turbulent medium. They perform a suite of simulations in which the initial turbulence is characterised either by purely compressive modes, or by a thermal mix of solenoidal and compressive modes (i.e. in the ratio 2:1), and they set the initial turbulent energy to be either subsonic or supersonic. For each case they perform ten simulations with different random seeds. Here we analyse the surface-density maps taken from these simulations to investigate how the fragmentation morphology appears on a $J$ plot.

They find that when the turbulence is subsonic, filaments fragment into quasi-periodic cores and the mix of turbulent modes (purely compressive or thermal) does not significantly influence the fragmentation. When the turbulence is supersonic, the filament first fragments into numerous sub-filaments. This change in structure can be identified by eye in the three dimensional volume-density cubes, as well as in two dimensional projected surface-density images. Using $J$ plots we can quantify this difference of morphology in an automated and objective way. 

Figure \ref{fig:filssubvsuper2D} shows the $J$ values of all the structures from the simulations of growing filaments in which the turbulence in the accreting material has a thermal mix of modes: (a) 10 simulations with subsonic turbulence, and (b) 10 simulations with supersonic turbulence. The dendrograms built on each surface-density image use the same parameters as \citet{2017Clarke+}, viz. min\_value $=10^{-20}\,{\rm g}\,{\rm cm}^{-3}$, min\_pixels $=100$ and min\_delta $=5\times 10^{-21}\,{\rm g}\,{\rm cm}^{-3}$. The simulations with {\it subsonic} turbulence (Figure \ref{fig:filssubvsuper2D}a) only create very long thin structures, similar in shape to the whole filament; the trunk of each dendrogram is shown as a black square in Figure \ref{fig:filssubvsuper2D}, and the other structures (branches and leaves) are simply the spine of the filament broken up into smaller structures. In contrast, the simulations with {\it supersonic} turbulence (Figure \ref{fig:filssubvsuper2D}b) create filamentary structures with a range of sizes and curvatures, filling more of the lower right quadrant on the $J$ plot. These are the curved sub-filaments described in \citet{2017Clarke+} and are significantly different in their morphology to the trunk structure which encompasses the whole filament.

\begin{figure}
\includegraphics[width=\linewidth]{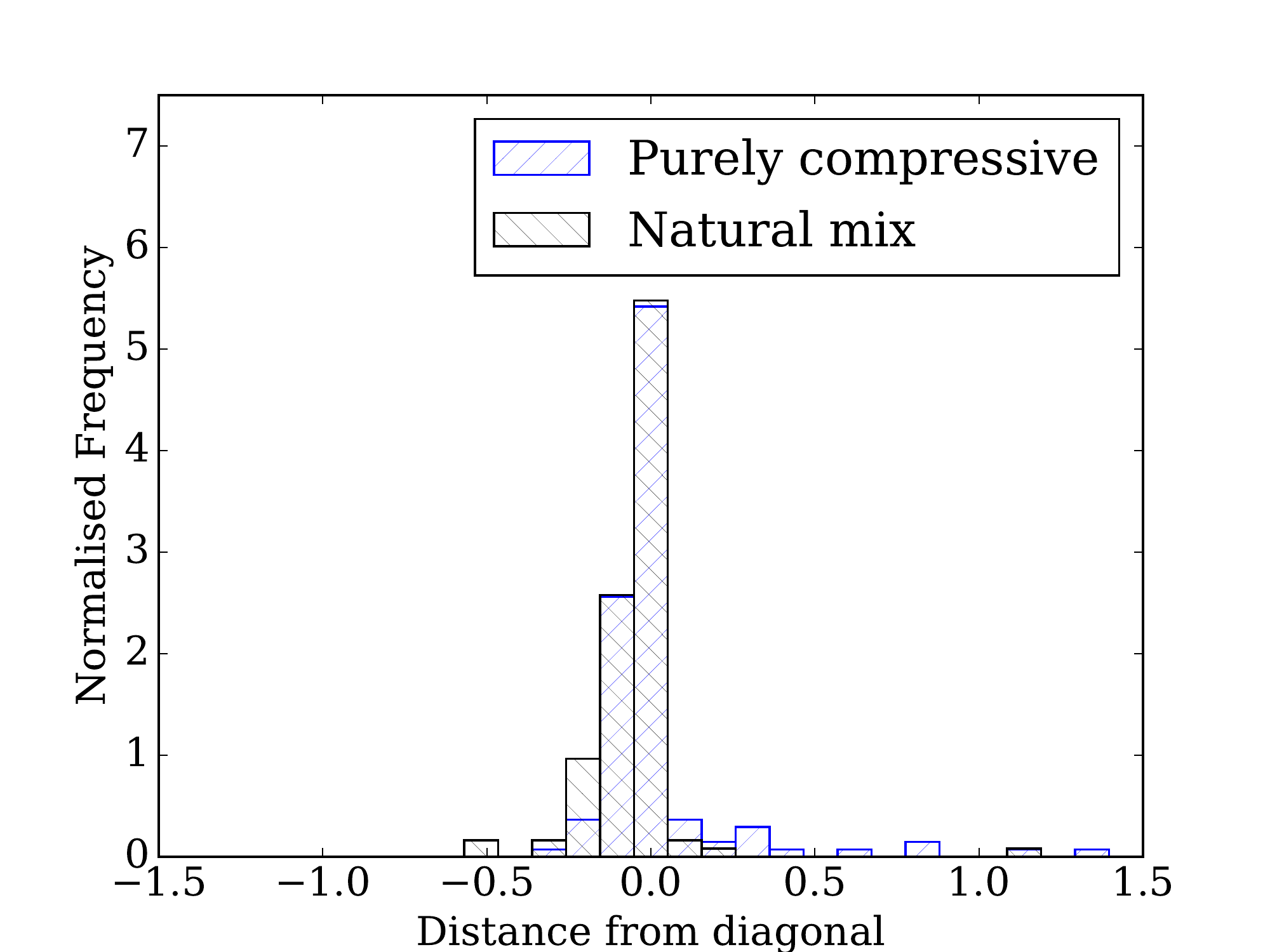}
\includegraphics[width=\linewidth]{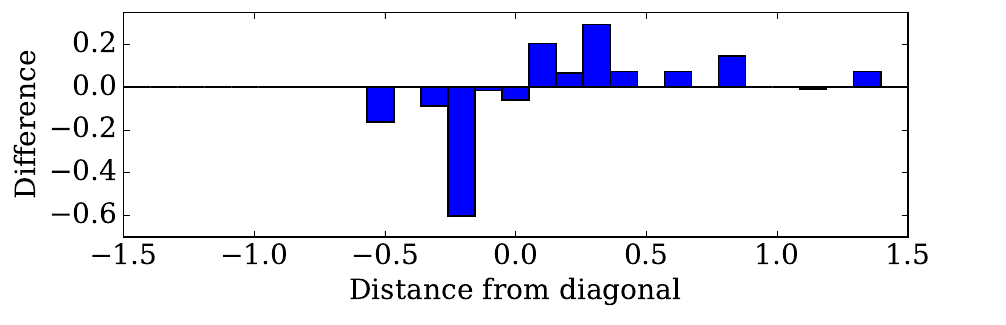}
\caption{Histograms showing the perpendicular distance from each point in the $J$ plot to the diagonal line $\Jo = -\Jt$. {\it Top.} The black striped histogram shows the results obtained with thermal mix of turbulent modes, and the blue striped histogram shows the results obtained with purely compressive turbulent modes, both analysed at 0.5 Myr. {\it Bottom.} The difference between the two histograms (purely compressive minus thermal mix in each bin). This highlights the excess of structures above and to the right of the diagonal in the purely compressive case (more centrally concentrated structures), and the excess below and to the left of the diagonal in the thermal-mix case (more curved filamentary structures).}
\label{fig:hist_diagonal_compnat}
\end{figure}

This change in morphology is seen predominantly in the simulations that invoke a thermal mix of solenoidal and compressive turbulence; when the turbulence is purely compressive, the formation of sub-filaments is suppressed. \citet{2017Clarke+} use dendrograms to quantify this. They find that in dendrograms built in three-dimensional PPP space using the volume-density distribution, the number of leaves in the dendrogram ($N_{_{\rm LEAVES}}$) is higher in the simulations using a thermal mix of turbulent modes (than those using purely compressive modes), indicating that the filaments are more sub-structured. The number of branches between the highest leaf and the trunk ($N_{_{\rm LEVELS}}$) is also higher in the simulations using a thermal mix of turbulent modes (than those using purely compressive modes), indicating that the sub-structure is more hierarchical. However, there is no significant difference in these measures if the dendrogram is built in two-dimensional PP space using surface-density images. $J$ plots therefore provide a new method for measuring morphological differences, using surface-density images.

Figure \ref{fig:filsnatvcomp2D} shows the $J$ values of all structures from the simulations of growing filaments in which the turbulence in the accreting material is supersonic: (a) 10 simulations with purely compressive modes, and (b) 10 simulations with a thermal mix of modes. In both cases, the main population of structures is filamentary, but in the case of purely compressive turbulent modes there are more centrally concentrated structures, while in the case of a thermal mix of turbulent modes there are more structures that are curved or even ring-like ($\Jo \leq$ 0). We can illustrate this difference more clearly by looking at the distance of each point from the diagonal line $\Jo = -\Jt$. This line represents the location in $J$ space of a theoretical `perfect' filament: an exactly symmetrical straight filament with uniform surface-density. Structures below and to the left of this line are more curved, and structures above and to the right of this line are more centrally concentrated. Figure \ref{fig:hist_diagonal_compnat} shows that, whilst both cases peak just below zero, the simulations with purely compressive turbulent modes have a significant positive tail of centrally concentrated structures, and the simulations with a thermal-mix of turbulent modes have a small negative tail indicating more curved structures. We can therefore use $J$ plots to identify objectively, in two dimensions, differences in structure that were previously identified only by eye or in three dimensions.

\subsubsection{Persistence of individual structures}

\begin{figure}
\centering
Subsonic \\
\includegraphics[width=\linewidth]{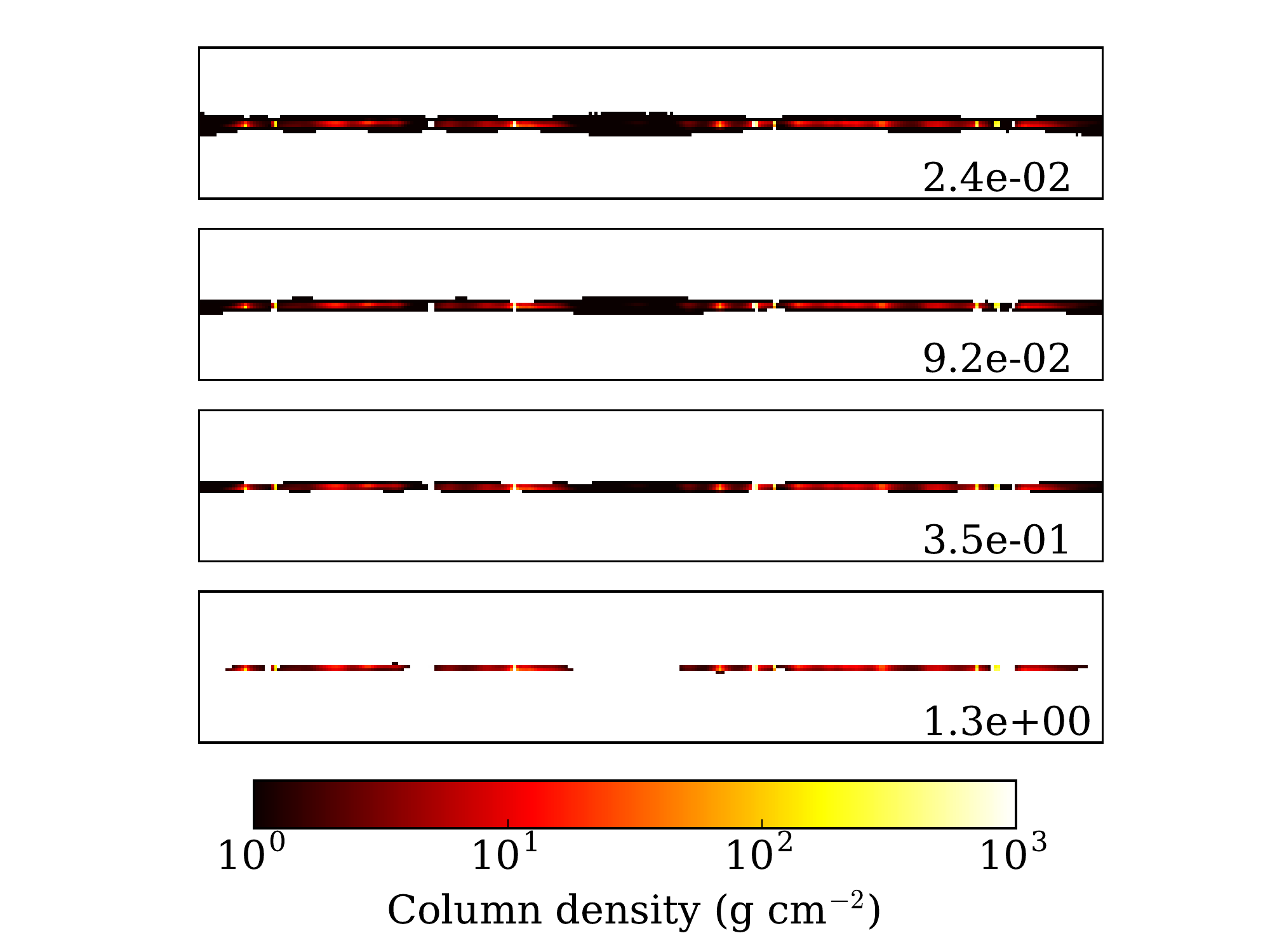}\\
Supersonic\\
\includegraphics[width=\linewidth]{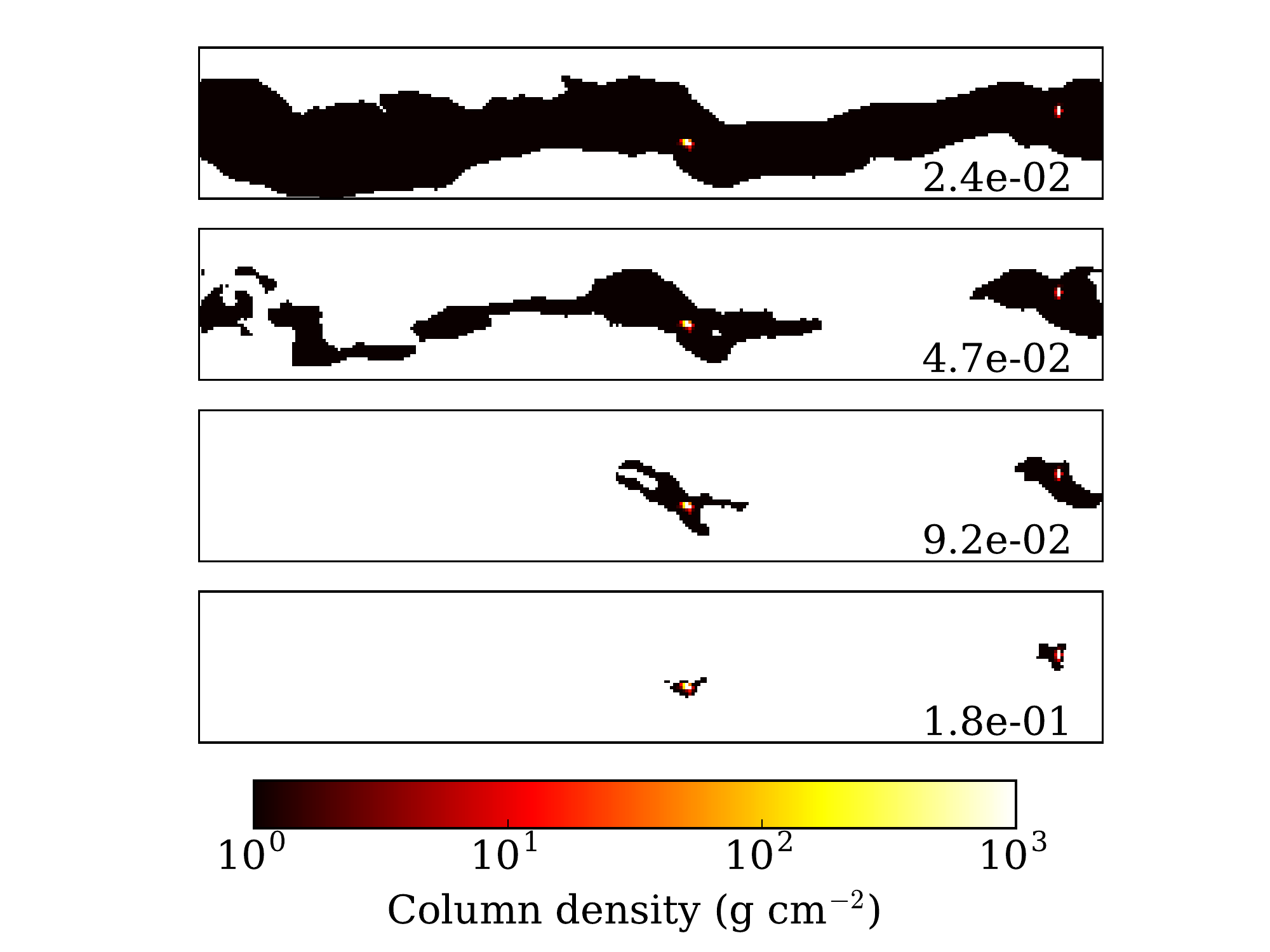}\\
\caption{Filaments analysed at different density thresholds. {\it Top:} subsonic turbulence. {\it Bottom:} supersonic turbulence. The surface-density threshold is given in the bottom right corner of each image. Only four of the ten surface-density thresholds analysed in Figure \ref{fig:singlestruct} are illustrated. For the subsonic turbulence case we show the first, third, fifth and seventh thresholds. For the supersonic case we show the first, second, third and fourth thresholds. The filaments are 3 pc in length and the scales are the same in all images.}
\label{fig:intensityfils}
\end{figure}

\begin{figure*}
\includegraphics[width=\linewidth]{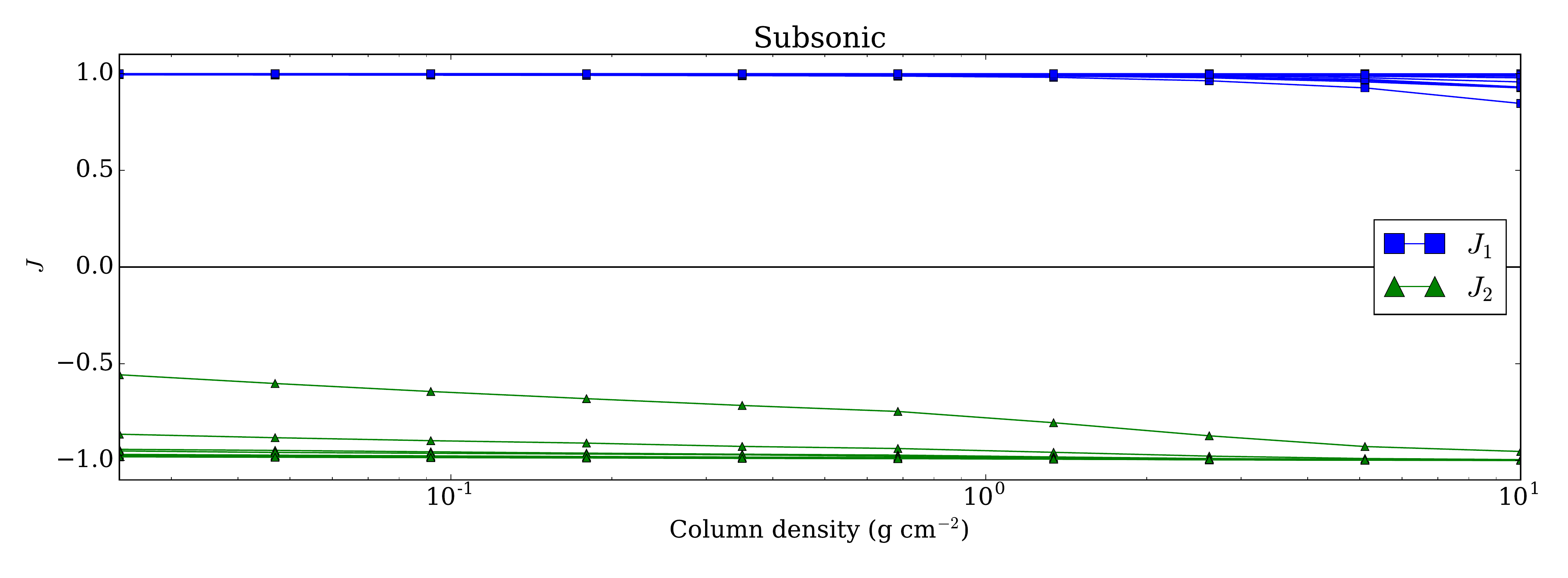}\\
\includegraphics[width=\linewidth]{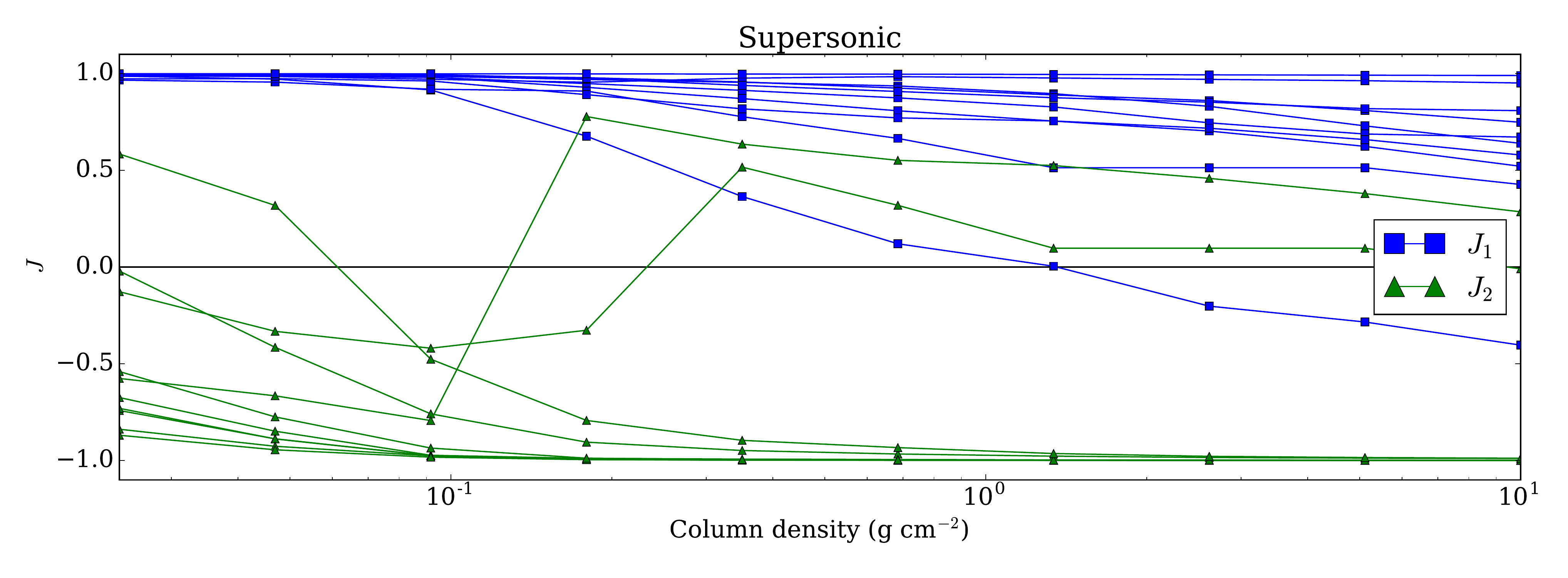}
\caption{The dependence of the $J$ moments on surface-density threshold for filaments forming in a medium with a thermal mix of turbulent modes. {\it Top:} 10 simulations with subsonic turbulence. {\it Bottom:} 10 simulations with supersonic turbulence. Filled blue squares represent values of $J_1$, and filled green triangles represent values of $J_2$.}
\label{fig:singlestruct}
\end{figure*}

Another way to characterise a dendrogram structure is to consider both the structure itself, and the smaller structures it contains -- for example a parent-branch and all the branches and leaves at higher surface-density thresholds that are descended from the parent-branch. We can then analyse how the $J$ moments change as the surface-density threshold is increased, from the minimum value at which the structure is first defined, to the maximum value in the densest leaf that it contains. In general, an increased $J$ moment means that the mass is more concentrated towards the centre of mass along the corresponding axis, and a decreased $J$ moment means that the mass is less concentrated towards the centre of mass along the corresponding axis. 

In Figure \ref{fig:intensityfils} we illustrate the results obtained for two typical filaments, one in which the turbulence in the accretion flow is subsonic, and one in which it is supersonic; in both cases there is a thermal mix of turbulent modes. In the subsonic case (top panel of Fig. \ref{fig:intensityfils}), the filament is very narrow, straight, and continuous; a few small gaps appear at the highest surface-density threshold. In the supersonic case (bottom  panel of Fig. \ref{fig:intensityfils}), the filament is much broader, bendier, and more fragmented; already, at a fairly low surface-density threshold, there are just two separate fragments.

Figure \ref{fig:singlestruct}a shows how the $J$ moments of the filaments in all ten {\it subsonic} cases evolve as the surface-density threshold is increased from $\sim 0.024\,{\rm g}\,{\rm cm}^{-2}$ to $\sim 1.3\,{\rm g}\,{\rm cm}^{-2}$.  With one exception, the $J$ moments remain very close to $(\Jo, \Jt) = (1,-1)$ at all thresholds, confirming objectively that the filaments are very straight and narrow, and the dense gas is rather evenly distributed along them. We describe this structure as `persistent', meaning that the $J$ moments do not change much when the surface-density threshold is increased. The one exception arises because the filament is slightly more fragmented in this particular case and so $J_2$ is somewhat higher. However, the pattern observed in the $J$ plots for the nine similar filaments is very specific, and might be used to identify automatically filaments forming in a medium where the turbulence is subsonic.

Figure \ref{fig:singlestruct}b shows how the $J$ moments of the filaments in all ten {\it supersonic} cases evolve as the surface-density threshold is increased from $\sim 0.024\,{\rm g}\,{\rm cm}^{-2}$ to $\sim 1.3\,{\rm g}\,{\rm cm}^{-2}$. For eight cases there is a clear pattern: at low thresholds, $J_1$ is of order unity, and $J_2$ is negative, as we would expect for filaments; then, at higher surface-densities, as the filament breaks up into two fragments, $J_1$ and $J_2$ decrease, with $J_2$ quickly tending to $\sim -1$. The two cases that do not conform to this pattern are ones in which the filament only produces a single fragment, and therefore, as the surface-density threshold increases, $J_2$ abruptly increases to positive values and then decreases towards zero. This is essentially a consequence of the fact that the simulation only treats a rather short filament. Most observed filaments contain more than one fragment, so the pattern observed in the $J$ plots for the eight similar filaments is probably more representative, and might be used to identify automatically filaments forming in a medium where the turbulence is supersonic.

\section{Conclusions}\label{SEC:Conclusions}%

We have developed a new method for analysing the structures in a segmented two dimensional image. Here we illustrate the method using dendrograms to segment the image into contiguous structures at varying surface-density thresholds. For each contiguous structure, the method first finds the area, $A$, mass, $M$, and principal moments of inertia, $I_1,I_2$. These parameters are then used to calculate the structure's $J$ moments (Eqn. \ref{EQN:JMoments}). The $J$ moments allow one to distinguish structures that are centrally concentrated from structures that are  centrally rarefied, and structures that are approximately circularly symmetric from structures that are elongated.

We apply the method to a tile from the Hi-GAL survey and show that it can identify and quantify the well-known RCW 120 bubble, as well as other ring-like structures in the same area. We also apply it to simulations of filaments growing in a turbulent medium and fragmenting, and show that $J$ plots are able to identify and quantify, objectively, the difference in projected structure that results from changes in the nature of the turbulence -- as measured by the mean Mach number of the turbulence, and the mix of compressive and solenoidal modes.

An analysis tool based on $J$ plots is freely available online at https://github.com/SJaffa/Jplots.

\section*{Acknowledgements}%

SEJ and ADPH gratefully acknowledge the support of postgraduate scholarships from the School of Physics \& Astronomy at Cardiff University and the UK Science and Technology Facilities Council. APW gratefully acknowledges the support of the consolidated grant ST/K00926/1 from the UK Science and Technology Facilities Council, and of the EU-funded {\sc vialactea} Network {\sc fp}7-{\sc space}-607380. SDC acknowledges support from the ERC starting grant No. 679852 `RADFEEDBACK'.

The computations were performed on the Cardiff University Advanced Research Computing facility, {\sc arcca}.This work uses observations made with the Herschel Space Observatory, a European Space Agency Cornerstone Mission with science instruments provided by European-led Principal Investigator consortia and significant participation from NASA. This work made use of \textsc{astrodendro}, a Python package to compute dendrograms of Astronomical data (http://www.dendrograms.org/), and Astropy, a community-developed core Python package for Astronomy (http://www.astropy.org; \citet{Astropy}).

\section*{Appendix. Computing principal moments and principal axes in 2D}\label{App1}

For a dendrogram structure represented by $P$ equal pixels, each with area $\Delta A$, we can compute the following moments,
\begin{eqnarray}\nonumber
M_0\!&\!=\!&\!\Delta A\sum\limits_{p=1}^{p=P}\left\{\Sigma_p\right\},\hspace{1.20cm}
M_x\;=\;\Delta A\sum\limits_{p=1}^{p=P}\left\{\Sigma_p x_p\right\},\\\nonumber
M_y\!&\!=\!&\!\Delta A\sum\limits_{p=1}^{p=P}\left\{\Sigma_p y_p\right\},\hspace{0.8cm}
M_{xx}\;=\;\Delta A\sum\limits_{p=1}^{p=P}\left\{\Sigma_p x_p^2\right\},\\\nonumber
M_{xy}\!&\!=\!&\!\Delta A\sum\limits_{p=1}^{p=P}\left\{\Sigma_p x_p y_p\right\},\hspace{0.5cm}
M_{yy}\;=\;\Delta A\sum\limits_{p=1}^{p=P}\left\{\Sigma_p y_p^2\right\},\\
\end{eqnarray}
where $\Sigma_p$ is the surface-density in pixel $p$, and $(x_p,y_p)$ is the position of its centre.

The centre of mass is then given by
\begin{eqnarray}
X&=&\frac{M_x}{M_0}\,,\hspace{0.5cm}Y\;\,=\;\,\frac{M_y}{M_0}\,,
\end{eqnarray}
and the moments about the Cartesian axes by
\begin{eqnarray}\nonumber
I_{xx}\!&\!=\!&\!M_{xx}-M_xX,\hspace{0.5cm}I_{xy}\;=\;M_{xy}-M_0XY,\\
I_{yy}&=&M_{yy}-M_yY.
\end{eqnarray}

The principal moments are then
\begin{eqnarray}\nonumber
I_{1,2}\!\!&\!\!=\!\!&\!\!\left(\frac{I_{xx}\!+\!I_{yy}}{2}\right)\mp\left\{\left(\frac{I_{xx}\!+\!I_{yy}}{2}\right)^2\!-\!\left(I_{xx}I_{yy}\!-\!I_{xy}^2\right)\right\}^{1/2},\\
\end{eqnarray}
(where the $\mp$ means that $I_1$ corresponds to the minus sign, and $I_2$ to the plus sign).

The first principal  axis has the equation
\begin{eqnarray}
y_1&=&Y\,+\,\frac{\left(I_1-I_{xx}\right)\,x}{I_{xy}}\,,
\end{eqnarray}
and the second principal axis
\begin{eqnarray}
y_2&=&Y\,-\,\frac{I_{xy}}{\left(I_1-I_{xx}\right)\,x}\,.
\end{eqnarray}

\bibliographystyle{mn2e}
\bibliography{papers}

\label{lastpage}

\end{document}